\documentclass[twocolumn,showpacs,aps,%
prb,amsmath,amssymb,floatfix,superscriptaddress]{revtex4}
\usepackage{amssymb,amsfonts}
\usepackage{epsf,epsfig}
\usepackage{graphicx} 
\usepackage{dcolumn}
\usepackage{bm}

\begin{document}

\title{Ground state fluctuations in rung-dimerized spin ladders}
\author{P.~N.~Bibikov}
\affiliation{Saint-Petersburg State University}
\date{\today}

\begin{abstract}
Treating an exactly rung-dimerized spin ladder as a reference model
we study perturbatively zero temperature quantum fluctuations in
spin ladders with slightly destroyed rung-dimerization. Analytic
expressions are obtained for the gas parameter (density of
rung-triplets) and the ground state energy per rung. At a strong
diagonal frustration as well as at a rather strong
antiferromagnetic rung coupling these results well agree with the
previous numerical calculations.
\end{abstract}
\pacs{75.10.Pq}

\maketitle

\section{Introduction}

Perturbation theory for rung-dimerized spin ladders first
considered in seminal works \cite{1,2} was then developed in a
number of papers \cite{3,4,5}. In all of them an array of isolated
rungs was treated as the exact zero-order reference model. The
main virtue of this approach is a possibility to perform calculations up
to a very high order. From the other hand its principal defect is
triviality of dispersion in excitation spectrum of the reference
(isolated rungs) model. According to this obstacle it seems
reliable that obtained results will be well
applicable only to narrow band spin ladders with $\Delta
E_{magn}\ll E_{gap}$ ($E_{gap}$ is a gap, $\Delta E_{magn}$ is a
magnon bandwidth). However this condition may be failed for real
compounds. For example a recent neutron scattering experiment
\cite{6} in the spin ladder compound $\rm(C_5D_{12}N)_2CuBr_4$
gives $\Delta E_{magn}/E_{gap}\approx0.72$ while for the spin
ladder compound $\rm La_6Ca_8Cu_{24}O_{41}$ this ratio is \cite{7}
about $5.3$.

In this paper we suggest a more general perturbative approach to magnetic excitations in spin ladders by treating the {\it exactly rung-dimerized} spin
ladder \cite{8,9,10} as a reference model . As in the isolated-rungs case \cite{1,2,3,4,5} the corresponding
ground state still is an array of rung-dimers (singlets)
\begin{equation}
|0\rangle_{r-d}=\prod_n|0\rangle_n,
\end{equation}
($|0\rangle_n$ is a singlet state associated with $n$-th rung),
however now an excited triplet (denoted by $|1\rangle_n$) can move
along the ladder from one rung to another. This process entails a
nontrivial dispersion for an elementary excitation
of the reference model. The latter will be
called a {\it rungon} because in fact it is an excited single rung
coherently propagating inside the rung-dimers bulk. It will be
implied that the perturbed model belongs to the {\it
rung-dimerized phase}, so its ground state is a dilute gas of
excited rungon pairs. In this picture an elementary excitation {\it
magnon} should be considered as a bare rungon interacting with the
perturbed ground state.

The main technical difficulty of the suggesting approach is {\it
non integrability} of the reference model \cite{10}. Indeed as it
will be shown in forward an evaluation of the lowest (second)
order perturbative correction to the ground state energy utilizes
only two-rungon wave functions \cite{9,10} of the reference model.
The next, i. e. one-magnon level requires a knowledge of the
three-rungon spectrum. Due to the non integrability this problem
is of a principal mathematical difficulty. Nevertheless for 5 {\it
integrable} cases the three-magnon problem may be solved by the
Bethe Ansatz \cite{10}.

In this paper the manifested approach is applied for an evaluation of
two main ground state parameters, namely the gas parameter (or
equivalently the density of rung triplets)
\begin{equation}
\rho=\lim_{N\rightarrow\infty}\frac{1}{N}\langle0|\hat Q|0\rangle.
\end{equation}
and the normalized energy density
\begin{equation}
\varepsilon=-\lim_{N\rightarrow\infty}\frac{\langle0|\hat
H|0\rangle}{NE_{rung}}.
\end{equation}
Here $\hat H$, $N$ and $E_{rung}$ are correspondingly the Hamiltonian,
a number of rungs and the excitation energy of an isolated rung. The
operator $\hat Q=\sum_nQ_n$, whose local density satisfies the
following relations
\begin{equation}
Q_n|0\rangle_n=0,\qquad Q_n|1\rangle_n=|1\rangle_n,
\end{equation}
is the number operator for rung-triplets. Probably it is better to
denote the gas parameter by $n$ (concentration) instead of $\rho$
(density). However in this paper the index $"n"$ is
utilized especially for enumerating of ladder rungs.

Both the quantities $\rho$ and $\varepsilon$ are of interest for
ascertaining of boundaries of the rung-dimerized phase
\cite{11,12}. Also $\rho$ may be considered as the governing
parameter for the dilute gas approximation \cite{13,14,15}. The
parameter $\varepsilon$ was used \cite{16} for evaluation of the spin
ladder entropy.

For a future development of the perturbation theory both $\rho$
and $\varepsilon$ may be considered as the corresponding {\it
governing} parameters. Indeed it is crucial to obtain a system of
conditions on the coupling constants under which the perturbative
approach is valid. In fact it should be a requirement for some
relevant dimensionless governing parameters to be small. Usually
only $\rho$ was suggested on this role. Namely the mean field
Green function approach to rung-dimerized spin ladders may be
developed \cite{14} only at $\rho\ll1$. The quasiclassical
approximation \cite{15} is valid only at $\rho\lambda(T)\ll 1$,
where $\lambda(T)$ is the average de Brogile wavelength of
thermally excited magnons expressed in lattice units. Despite the
parameter $\varepsilon$ was not yet utilized for establishing of
the perturbative approach validity we suppose that this {\it
energetic} quantity also is very important.

In this light it seems reasonable that besides numerical
estimations \cite{11,12} it is necessary to have for $\rho$ and
$\varepsilon$ reliable analytic formulas. The latter are just
obtained in the present paper in the second order of perturbation
theory near the exact rung-dimerized ground state (1). As it was
already mentioned technically these calculations are possible
because the first order correction term in the perturbative expansion for $|0\rangle$ contains
only (previously obtained by the author \cite{9,10}) two-rungon
excitations around $|0\rangle_{r-d}$. At $\rho\ll1$ and
$\varepsilon\ll1$ the obtained formulas agree with the
corresponding numerical data \cite{11}.

\section{Structure of the Hamiltonian}
It is more convenient to represent the general spin ladder
Hamiltonian in the following form \cite{10}
\begin{equation}
\hat H=\hat H_{r-d}+J_6\hat V,
\end{equation}
where
\begin{eqnarray}
\hat H_{r-d}&=&\sum_nJ_1(Q_n+Q_{n+1})+
J_2({\bf\Psi}_n\cdot\bar{\bf\Psi}_{n+1}+\bar{\bf\Psi}_n\cdot{\bf\Psi}_{n+1})\nonumber\\
&+&J_3Q_nQ_{n+1}+J_4{\bf S}_n\cdot{\bf S}_{n+1}+J_5({\bf
S}_n\cdot{\bf S}_{n+1})^2,
\end{eqnarray}
and
\begin{equation}
\hat
V=\sum_n\bar{\bf\Psi}_n\cdot\bar{\bf\Psi}_{n+1}+{\bf\Psi}_n\cdot{\bf\Psi}_{n+1}.
\end{equation}
Here
\begin{eqnarray}
{\bf\Psi}_n&=&\frac{1}{2}\big({\bf S}_{1,n}-{\bf
S}_{2,n}\big)-i{[}{\bf S}_{1,n}\times{\bf
S}_{2,n}{]},\nonumber\\
{\bf\bar\Psi}_n&=&\frac{1}{2}\big({\bf S}_{1,n}-{\bf
S}_{2,n}\big)+i{[}{\bf S}_{1,n}\times{\bf S}_{2,n}{]},\\
{\bf S}_n&=&{\bf S}_{1,n}+{\bf S}_{2,n}\nonumber\\
Q_n&=&\frac{1}{2}{\bf S}_n^2={\bf\bar\Psi}_n\cdot{\mathbf\Psi}_n,
\end{eqnarray}
(${\bf S}_{i,n}$ ($i=1,2$) are spin-1/2 operators associated with
$n$-th rung). According to
\begin{eqnarray}
{\bf\bar\Psi}^a_n|0\rangle_n=|1\rangle^a_n,&\quad&{\bf\bar\Psi}^a_n|1\rangle^b_n=0,\nonumber\\
{\bf\Psi}^a_n|0\rangle_n=0,&\quad&
{\bf\Psi}^a_n|1\rangle^b_n=\delta_{ab}|0\rangle_n,
\end{eqnarray}
the operators ${\bf\bar\Psi}$ and ${\bf\Psi}$ may be considered as
(neither Bose, nor Fermi) creation-annihilation operators for
rung-triplets. Correspondence between (5)-(7) and the traditional
representation
\begin{eqnarray}
\hat
H&=&\sum_{n=-\infty}^{\infty}H^r_{n,n+1}+H^l_{n,n+1}+H^d_{n,n+1}\nonumber\\
&+&H^{rr}_{n,n+1}+H^{ll}_{n,n+1}+H^{dd}_{n,n+1} +J_{norm},
\end{eqnarray}
where
\begin{eqnarray}
H^r_{n,n+1}&=&
\frac{J_r}{2}({\bf S}_{1,n}\cdot{\bf S}_{2,n}+{\bf S}_{1,n+1}\cdot{\bf S}_{2,n+1}),\nonumber\\
H^l_{n,n+1}&=& J_l({\bf S}_{1,n}\cdot{\bf S}_{1,n+1}+{\bf
S}_{2,n}\cdot{\bf
S}_{2,n+1}),\nonumber\\
H^d_{n,n+1}&=&J_d({\bf S}_{1,n}\cdot{\bf S}_{2,n+1}+{\bf
S}_{2,n}\cdot{\bf
S}_{1,n+1}),\nonumber\\
H^{rr}_{n,n+1}&=&J_{rr}({\bf S}_{1,n}\cdot{\bf
S}_{2,n})({\bf S}_{1,n+1}\cdot{\bf S}_{2,n+1}),\nonumber\\
H^{ll}_{n,n+1}&=&J_{ll}({\bf S}_{1,n}\cdot{\bf S}_{1,n+1})({\bf
S}_{2,n}\cdot{\bf S}_{2,n+1})\nonumber\\
H^{dd}_{n,n+1}&=&J_{dd}({\bf S}_{1,n}\cdot{\bf S}_{2,n+1})({\bf
S}_{2,n}\cdot{\bf S}_{1,n+1}),\nonumber\\
J_{norm}&=&\frac{3}{4}\Big(J_r+J_l-J_d\Big)-\frac{9}{16}J_{rr}-\frac{3}{8}J_{ll},
\end{eqnarray}
is given by the formulas \cite{10}
\begin{eqnarray}
J_1&=&\frac{1}{4}\Big(2J_r-3J_{rr}-J_{ll}-J_{dd}\Big),\nonumber\\
J_2&=&\frac{1}{8}\Big(4(J_l-J_d)+J_{ll}-J_{dd}\Big),\nonumber\\
J_3&=&J_{rr},\nonumber\\
J_4&=&\frac{1}{8}\Big(4(J_l+J_d)+J_{ll}+J_{dd}\Big),\nonumber\\
J_5&=&\frac{1}{4}\Big(J_{ll}+J_{dd}\Big),\nonumber\\
J_6&=&\frac{1}{8}\Big(4(J_l-J_d)-J_{ll}+J_{dd}\Big).
\end{eqnarray}

Let us notice that one can inverse the relations (8) and express
the spin operators
\begin{eqnarray}
{\mathbf
S}_{1,n}&=&\frac{1}{2}\Big({\bf\Psi}_n+{\bf\bar\Psi}_n-i[{\bf\bar\Psi}_n\times{\bf\Psi_n}]\Big),\nonumber\\
{\mathbf
S}_{2,n}&=&\frac{1}{2}\Big(-{\bf\Psi}_n-{\bf\bar\Psi}_n-i[{\bf\bar\Psi}_n\times{\bf\Psi_n}]\Big).
\end{eqnarray}
As it was already mentioned \cite{10} the operators (8) do not coincide
with very similar ones introduced in Refs. 3 and 17,18 (as well as
in other papers of these authors). Indeed the representation (14)
is similar to the ones suggested in Refs. 17 and 3, but in fact is
not identical to any of them because the analogs of ${\bf\Psi}_n$
and ${\bf\bar\Psi}_n$ treated in these Refs. act in extended
vector spaces. That is why for example the "inverse" formula (8)
fails for these operators. The operators suggested in Ref. 18 in
fact act in the same vector space as (8) but have a slightly
different form.

According to (6)
\begin{eqnarray}
E_{rung}&=&2J_1,\\
\rho&=&-\frac{\partial}{\partial J_1}\Big(J_1\varepsilon\Big).
\end{eqnarray}
Without loss of generality one can imply $J_2>0$ because a
substitution $J_2\rightarrow-J_2$ is equivalent to permutation of
spins in all odd (or even) rungs.

As it readily follows from (4), (6) and (10) ${[}\hat Q,\hat
H_{r-d}{]}=0$. So the Hilbert space related to $\hat H_{r-d}$
splits into a direct sum
\begin{equation}
{\cal H}=\sum_{m=0}^{\infty}{\cal H}^m,\quad \hat Q|_{{\cal
H}^m}=m,
\end{equation}
where each ${\cal H}^m$ corresponds to $m$-rungon sector.

For rather big $J_1$ vector (1) is the (zero energy) ground state
of $\hat H_{r-d}$. The complete set of inequalities on $J_1$ which
guarantees the exact rung-dimerization is not yet obtained except
the following one \cite{10}
\begin{equation}
J_1-J_2>0.
\end{equation}

According to (7) and (10)
\begin{equation}
\hat V:\quad {\cal H}^m\rightarrow{\cal H}^{m-2}\oplus{\cal
H}^{m-2}.
\end{equation}
So the operator $\hat V$ destroys the exact rung-dimerization.
However, it seems reliable that the region $J_1>J_{2,3,4,5}>J_6$
should contain the rung-dimerized phase for which the operator
$J_6\hat V$ may be treated perturbatively and
\begin{equation}
\rho\ll1,\quad\varepsilon\ll1.
\end{equation}

The parameter $J_1$ characterizes a chemical potential of an
exited rung (or rungon mass) while $J_2$ its kinetic energy. The
couplings $J_3-J_5$ describe a spin-dependent rungon-rungon
interaction. Finitely the parameter $J_6$ governs
creation-annihilation of singlet rungon pairs.

\section{Evaluation of the governing parameters}

According to Eqs. (3) and (19) $_{r-d}\langle0|\hat
V|0\rangle_{r-d}=0$, so
\begin{equation}
\varepsilon=\varepsilon_{bound}+\varepsilon_{scatt}+o(J_6^2),
\end{equation}
where
\begin{eqnarray}
\varepsilon_{bound}&=&\frac{J_6^2}{2J_1}\lim_{N\rightarrow\infty}\frac{|\langle
bound|\hat V|0\rangle_{r-d}|^2}{NE_{bound}},\nonumber\\
\varepsilon_{scatt}&=&\frac{J_6^2}{2\pi
J_1}\int_0^{\pi}\frac{|\langle q,scatt|\hat
V|0\rangle_{r-d}|^2}{E_{scatt}(q)}dq
\end{eqnarray}
(we have utilized
$1/N\sum_q\rightarrow1/(2\pi)\int_0^{2\pi}d(2q)$) are
contributions from singlet, translationary invariant
bound and scattering two-rungon states \cite{9,10}
\begin{eqnarray}
|bound\rangle&=&\sqrt{\frac{\Delta_0^2-1}{3N}}\sum_{m<n}
\Delta_0^{m-n} ...|1\rangle^j_m...|1\rangle^j_n...,\nonumber\\
|q,scatt\rangle&=&\frac{\sqrt{2}}{\sqrt{3(\Delta^2_0-2\Delta_0\cos{q}+1)}N}\nonumber\\
&\times&\sum_{m<n} a(q,n-m) ...|1\rangle^j_m...|1\rangle^j_n...
\end{eqnarray}
Here $0<q<\pi$ is a half of relative wave number
\begin{equation}
a(q,n)=\sin{nq}-\Delta_0\sin{(n-1)q},
\end{equation}
and $\Delta_0=(J_3-2J_4+4J_5)/(2J_2)$.

The corresponding energies are the following
\begin{eqnarray}
E_{bound}&=&4J_1+2J_2\Big(\Delta_0+\frac{1}{\Delta_0}\Big)\nonumber\\
&=&2E_{gap}+\frac{(\Delta_0+1)^2}{2\Delta_0}\Delta
E_{rung},\\
E_{scatt}(q)&=&4(J_1+J_2\cos{q}).
\end{eqnarray}
where $E_{gap}=2(J_1-J_2)$ and $\Delta E_{rung}=4J_2$ are the
rungon gap and energy width \cite{10}.

As it follows from (23) the translationary invariant bound state
exists only for
\begin{equation}
|\Delta_0|>1.
\end{equation}
(however as it follows from (25) and (26) in a real compound it
should metastable at $\Delta_0>1$).

According to (7), (10) and (23)
\begin{eqnarray}
\langle q,scatt|\hat V|0\rangle_{r-d}
&=&\frac{\sqrt{6}\sin{q}}{\sqrt{(\Delta^2_0-2\Delta_0\cos{q}+1)}},\nonumber\\
\langle bound|\hat
V|0\rangle_{r-d}&=&\sqrt{N}\frac{\sqrt{3(\Delta^2_0-1)}}{\Delta_0}.
\end{eqnarray}

Substituting (28) into (22) and utilizing (25)-(27) one can
readily obtain
\begin{eqnarray}
\varepsilon_{scatt}&=&\frac{3J_6^2}{8J_1\Delta_0}{\cal I},\nonumber\\
\varepsilon_{bound}&=&\Theta(\Delta_0^2-1)\frac{3J_6^2(\Delta_0^2-1)}
{2J_1\Delta_0^2E_{bound}}.
\end{eqnarray}
($\Theta(x)=1$ for $x>0$ and $\Theta(x)=0$ for $x<0$), where
\begin{equation}
{\cal I}=\frac{1}{2\pi i
}\oint_{|z|=1}\frac{(z^2-1)^2dz}{z(z-\Delta_0)(z-1/\Delta_0)(J_2z^2+2J_1z+J_2)}.
\end{equation}

For calculation of $\cal I$ we note that according to (18) the
pole $z=-(J_1+\sqrt{J_1^2-J_2^2})/J_2$ lies outside the unit
circle. Therefore only residues in the poles $z=0$,
$z=(\sqrt{J_1^2-J_2^2}-J_1)/J_2$ and one of the appropriate points
$1/\Delta_0$ (for $|\Delta_0|>1$) or $\Delta_0$ (for
$|\Delta_0|<1$) give contributions to $\cal I$. Performing the
calculations one can obtain
\begin{eqnarray}
\varepsilon_{scatt}&=&\frac{3J_6^2}{8J_1J_2\Delta_0}\Big(1\nonumber\\
&-&\frac{J_2|\Delta_0^2-1|+2\Delta_0\sqrt{J_1^2-J_2^2}}{
[2\Delta_0J_1+(\Delta_0^2+1)J_2]}\Big).
\end{eqnarray}

Then according to (16)
\begin{equation}
\rho=\rho_{bound}+\rho_{scatt}+o(J_6^2),
\end{equation}
where
\begin{eqnarray}
&\rho_{scatt}=\frac{\displaystyle3J_6^2}{\displaystyle4[2\Delta_0J_1+(\Delta_0^2+1)J_2]^2}&\nonumber\\
&\times\Big[\frac{\displaystyle(\Delta_0^2+1)J_1+2\Delta_0J_2}{\displaystyle\sqrt{J_1^2-J_2^2}}-|\Delta_0^2-1|\Big],&
\nonumber\\
&\rho_{bound}=6\Theta(\Delta_0^2-1)\frac{\displaystyle(\Delta_0^2-1)J_6^2}{\displaystyle\Delta_0^2E_{bound}^2}.&
\end{eqnarray}

The denominators $\sqrt{J_1^2-J_2^2}\propto\sqrt{E_{gap}}$ and
$E_{bound}^2$ in (33) describe the ground state
(1) destruction caused by condensation of scattering and bound singlet pairs.

\section{Comparison with numerical results}

Here we compare the DMRG data \cite{11} for the case
\begin{equation}
0\leq J_r\leq 2,\quad J_l=1,\quad J_{rr}=J_{ll}=J_{dd}=0,
\end{equation}
(also in Ref. 11 was taken $J_{norm}=0$, so we had to shift the data for $varepsilon$) with the corresponding results
obtained analytically with the use of the perturbative formulas
(21), (29), (31) and (32), (33)  It is more instructive to do it
in the case of maximal dimerization (considered in Ref. 11) namely
for $J_r=2$.

For $J_d=0$ the Ref. 11 gives $\rho=0.11$, $\varepsilon=0.11$ (our
results: $\rho=0.22$, $\varepsilon=0.14$). For $J_d=0.2$,
$\rho=0.10$ and $\varepsilon=0.08$ (our results: $\rho=0.14$,
$\varepsilon=0.09$). For $J_d=0.4$, $\rho=0.07$ and $\varepsilon
=0.05$ (our results: $\rho=0.09$, $\varepsilon=0.05$). For
$J_d=0.6$, $\rho=0.04$ and $\varepsilon=0.03$ (our results:
$\rho=0.04$, $\varepsilon=0.03$). For $J_d=0.8$, $\rho=0.01$ and
$\varepsilon<0.01$ (our results: $\rho=0.01$,
$\varepsilon=0.007$). Comparing numerical and analytic results one
may conclude that the suggested perturbative approach works better
for frustrated ladders. Really at $J_d=0.8$ even for $J_r=1.6$ the
numerical \cite{11} result $\rho=0.03$ coincides with the
theoretical one. Also for $J_d=0.8$ and $J_r=1.5$ in the both
approaches $\varepsilon=0.02$.

A discrepancy between the numerical data and analytical
predictions for non frustrated spin ladders indicates that besides
dynamics of single rungons their properties are governed by
creation-annihilation processes. However the latter may be
suppressed by strong antiferromagnetic rung coupling.

\section{Summary}

In this paper the perturbative formulas for the gas
parameter (Eqs. (32), (33)) and ground state energy energy per
rung (Eqs. (21), (29), (31)) were obtained for spin ladders
belonging to the rung-dimerized phase. At strong diagonal
frustration as well as at rather strong antiferromagnetic rung
coupling the result agrees with the previous DMRG calculations. In this region
the conditions (20) are satisfied so one may conclude that the system really
lies in the rung-dimerized phase.

It will be interesting to compare the analytical and numerical
approaches more precisely and in a wider range of the
coupling parameters including four-spin terms.

The author is grateful to M.~I. Vyazovsky, S.~V. Maleyev, S.~L.
Ginzburg and A.~V. Syromyatnikov for helpful discussions.

\end{document}